# Tachyonic dark matter

P.C.W. Davies
Australian Centre for Astrobiology
Macquarie University, New South Wales, Australia 2109

pdavies@els.mq.edu.au

**Abstract**

Recent attempts to explain the dark matter and energy content of the universe have involved some radical extensions of standard physics, including quintessence, phantom energy, additional space dimensions and variations in the speed of light. In this paper I consider the possibility that some dark matter might be in the form of tachyons. I show that, subject to some reasonable assumptions, a tachyonic cosmological fluid would produce distinctive effects, such as a surge in quantum vacuum energy and particle creation, and a change in the conventional temperature-time relation for the normal cosmological material. Possible observational consequences are discussed.

## 1. Tachyons in an expanding universe

In this section I consider the behaviour of a tachyon in an expanding universe, following the treatment in Davies (1975). A tachyon is a particle with imaginary mass $i\mu$ ($\mu$ real and positive), velocity $v > c$ and momentum and energy given in a local inertial frame by

$$p = \mu v (v^2 - 1)^{-1/2} \qquad (1.1)$$

$$E = \mu (v^2 - 1)^{-1/2} \qquad (1.2)$$

where here and henceforth I choose units with $c = \hbar = 1$. Consider such a particle moving in a Friedmann-Roberston-Walker (FRW) universe with scale factor $a(t)$, $t$ being the cosmic time. In a short time $dt$, the particle will have moved a distance $vdt$ to a point where the local comoving frame is retreating at a speed $dv = (a'/a)vdt$, where $a' = da/dt$. In this new frame, the tachyon will now have a velocity

$$v + dv = [v - (a'/a)vdt]/[1 - v^2(a'/a)dt]. \qquad (1.3)$$

Note that because $v > 1$ the velocity of the tachyon in the new comoving frame is actually *greater*, i.e. the expansion of the universe causes the tachyon to speed up, rather than slow down, unlike the case with conventional particles of non-zero rest mass (tardyons).
Equation (1.3) may be integrated to obtain

$$v = (1 - a^2/A^2)^{1/2} \qquad (1.4)$$

where $A$ is a constant of integration. The energy of the tachyon is given by Eq. (1.2):

$$E = \mu(A^2/a^2 - 1)^{1/2} \qquad (1.5)$$

which falls as the universe expands, as expected. Results (1.4) and (1.5) may be derived more formally by integrating the geodesic equation for a tachyon (Davies 1975).
The constant of integration $A$ may be fixed by selecting an initial velocity or energy for the tachyon. Imagine a homogeneous isotropic tachyon cosmological fluid. For simplicity I shall assume the tachyons do not interact with each other or with tardyons, but for consistency they must interact with gravitons. Strong coupling to gravitons would have occurred only in the very early universe, and one may posit that the tachyonic and graviton fluids would have been in thermodynamic equilibrium. A plausible initial condition is therefore $E = E_{\text{Planck}}$ when $a = a_{\text{Planck}}$. For these high energies Eq. (1.5) approximates to $\mu A/a$, which fixes

$$A = E_{\text{Planck}} a_{\text{Planck}}/\mu. \qquad (1.6)$$

Another case of interest is that the tachyons are created with GUT energies during the re-heating of the universe at the end of an inflationary era. In that case

$$A = E_{\text{GUT}} a_{\text{reheating}}/\mu. \qquad (1.7)$$

From Eqs. (1.4) and (1.5) it is apparent that when $a \to A$ the velocity of the tachyon approaches infinity and the energy tends to zero. In effect, the tachyon disappears from the universe. This may be interpreted (Davies 1975) as a tachyon-antitachyon annihilation event with zero energy released. When will this disappearing act occur? A rough estimate may be made by treating the universe as radiation-dominated, in which case $a(t) \propto t^{1/2}$, and Eqs. (1.6) and (1.7) yield, in both cases,

$$t_{\text{disappearance}} \approx 10^{13} \text{s}/[\mu(\text{eV})]^2. \tag{1.8}$$

Of course, we have no idea what the imaginary mass of the tachyon might be. If $\mu \sim$ proton mass, $t_{\text{disappearance}} \sim 10^{-5}$ s. At the other extreme, if $\mu \sim$ neutrino mass, perhaps as low as 1 eV, then $t_{\text{disappearance}} \sim 10^6$ years, i.e. the tachyon disappearance takes place during the cosmic 'dark age' after photon decoupling and before star formation.

If tachyons were to form a component of dark matter at the current epoch, it is clear that $\mu$ would have to be exceedingly small, say $< 10^{-3}$ eV. But from (1.2) we see that $E < \mu c^2$, so the total energy density of tachyons would then be

$$\rho_{\text{tachyons}} < n\mu \tag{1.9}$$

where $n$ is the number density of tachyons. If there are $N$ distinct species of tachyons with sufficiently low $\mu$ in the cosmological fluid, and all were in thermodynamic equilibrium with the universe at the Planck time, then $n \sim N \times$ number density of gravitons, and

$$\rho_{\text{tachyons}} < N\rho_{\text{gravitons}} < N\rho_{\text{photons}} \tag{1.10}$$

as the graviton background is suppressed relative to the photon background (Weinberg 1972). Obviously $\rho_{\text{tachyons}}$ is negligible unless $N$ is a very large number, say $> 10^6$, which seems unreasonable. So we may conclude that tachyons will constitute a negligible fraction of the dark matter at this epoch, unless either (i) $N$ is much larger than for tardyons or, (ii) $n$ is determined not by a condition of early thermodynamic equilibrium, but by some other criterion. Because tachyons enter our region of the universe on spacelike trajectories, they are not subject to the normal cosmological initial conditions, so we are free to posit alternatives to the ansatz (1.6) or (1.7). In the absence of a theory of tachyon formation, however, there is nothing much we can say about these alternatives.

## 2. Cosmological dynamics

The presence of a tachyon fluid will modify the dynamics of the FRW model. For simplicity I shall restrict attention to the $k = 0$ (spatially flat) case. The scale factor $a(t)$ satisfies the FRW equations

$$a'^2/a^2 = 8\pi G\rho \tag{2.1}$$

$$2a''/a + (a'/a)^2 = -8\pi Gp. \tag{2.2}$$

Assuming for the moment that the tachyons all have the same energy at the same cosmic time t in their local comoving frame, the energy density will be given from Eq. (1.5) by

$$\rho_{\text{tachyons}} = n\mu(A^2/a^2 - 1)^{1/2}. \tag{2.3}$$

If the universe contained only tachyons, then Eqs. (2.1) and (2.3) yield

$$Ct = 1 - (1 - a^2/A^2)^{3/4} \tag{2.4}$$

where C is a constant of integration determined by the total energy density. For small $a$, the universe resembles a radiation-filled FRW model with $a \propto t^{1/2}$. But at $a = A$, $t = 1/C$, the tachyons disappear, and the universe becomes a patch of Minkowski space thereafter. This discontinuity is less abrupt than might appear at first sight, because $\rho_{\text{tachyons}} \to 0$ there; indeed, from Eq. (2.1) or (2.3) one sees that $a'/a \to 0$. However, Eq. (2.4) reveals an infinite discontinuity in $a''$ at the vanishing point $a = A$:

$$a''/a \to -(C/3)(1 - a^2/A^2)^{-3/4}, \quad t < 1/C$$

$$= 0 \qquad\qquad t > 1/C. \tag{2.5}$$

It follows from Eq. (2.2) that the divergence in $a''/a$ at $a = A$ corresponds to an infinite momentary pressure when the tachyons disappear. The source of this divergence can be traced to the behaviour of the tachyons' momentum as the zero-energy condition is approached. From Eq. (1.1) it is clear that p remains finite ($\to \mu$) as $v \to \infty$ and $E \to 0$. Thus there is an infinite flux of momentum. One may also deduce the same result from the conservation of energy condition

$$p\,da^3 + d(\rho a^3) = 0 \tag{2.6}$$

using Eq. (2.3). This behaviour also implies a discontinuity in the scalar curvature R, given here by

$$R = 6(a'^2/a^2 + a''/a). \tag{2.7}$$

In a more realistic case the cosmological fluid will have both tachyonic and non-tachyonic components. If the initial energy density ratio is

$$\rho_{\text{tachyons}}/\rho_{\text{radiation}} = \alpha \tag{2.8}$$

then in place of Eq. (2.5) one obtains

$$Ct = [2 - \alpha(1 - a^2/A^2)^{1/2}][1 + \alpha(1 - a^2/A^2)^{1/2}]^{1/2} - (2 - \alpha)(1 + \alpha)^{1/2} \qquad a < A$$

$$= (3\alpha/4)(a^2/A^2 - 1) + 2 - (2 - \alpha)(1 + \alpha)^{1/2} \qquad\qquad a > A. \tag{2.9}$$

The same general features are found: $a \propto t^{1/2}$ at early times, followed by a deviation from the standard FRW behaviour, culminating in an abrupt disappearance of the tachyons at $a = A$, standard FRW behaviour thereafter, and an infinite discontinuity in $a''/a$.

The assumption that all the tachyons have the same energy at any given $t$ is obviously unrealistic. If the tachyonic fluid were in thermodynamic equilibrium with the universe at the Planck time, the energies would be distributed with a thermal spectrum. This translates into a distribution of values of $A$, which has the effect of smearing the infinite discontinuity in $a''/a$ and $R$. Nevertheless, there will still be a large increase in $a''/a$ and $R$, which prompts the question of whether or not this will produce observable effects.

**Quantum flash**

The abrupt behaviour in $a''/a$ at $a = A$ may not be important for the cosmological dynamics (the Hubble constant remains continuous there), but it does have an effect on the state of the quantum vacuum. In particular, it can produce a burst of particle creation. To illustrate this phenomenon, I consider a massless scalar field $\varphi$ propagating in a $k = 0$ FRW universe. Suppose the field is initially in the conformal vacuum state (Birrell & Davies 1982). Using first order perturbation theory, the number density of created particles is given by (Birrell & Davies 1982, Eq. (5.118))

$$n_\varphi = (9/4\pi a^3)(\xi - 1/6)^2 \int (aa'^2 + a^2 a'')^2 \, dt \qquad (3.1)$$

where $\xi$ is the conformal coupling parameter, and $t$ runs over a range of values from the 'in' region where the vacuum is defined to the 'out' region where the particles are examined. Using the solution (2.4) for the scale factor, substituting $da/a'$ for $dt$ and choosing a range of integration that includes the instant of tachyon disappearance at $a = A$, one finds that the integral in Eq. (3.1) diverges like $(1 - a^2/A^2)^{-1/4}$ as $a \to A$. This result suggests that the tachyon disappearance induces dramatic effects in the quantum vacuum of the scalar field – a 'quantum flash' – the formal divergence being quenched by the smearing in A values implicit in assuming a thermal distribution of tachyons.

Before drawing general conclusions, however, three points must be made. First, the use of perturbation theory is obviously inconsistent with $a'' \to \infty$, so Eq. (3.1) may understate the particle creation effect. The extent to which this is the case will depend on the assumptions made about the smearing of $A$ values required by demanding a distribution of initial tachyon energies. Second, if the parameter $\xi$ is chosen to make the scalar field conformally invariant, corresponding most closely to, say, the electromagnetic field, then $\xi = 1/6$ and the right hand side of Eq. (3.1) vanishes. So the burst of particle production requires the breaking of conformal symmetry. In a more realistic model, conformal symmetry might be broken even when $\xi = 1/6$ by taking into account departures from homogeneity and isotropy in the disappearance of the tachyons. But this is a different and more complicated calculation.

Finally, the strength of the divergence in the integral in Eq. (3.1) is due in part to the artificiality of the model based on Eq. (2.4), which assumes a universe containing only tachyons. This requires that $a' = 0$ at $a = A$, which introduces a factor $(1 - a^2/A^2)^{-1/4}$ in the integrand arising from the substitution $dt = da/a'$. If normal matter is also present, then $a' \neq 0$ at $a = A$, and the divergence will be only logarithmic. This may be verified for the more realistic model solution (2.9) where, to leading order

$$n_\phi = (C^3/a^3)(\xi - 1/6)^2[\beta + \gamma \ln(1 - a/A)] \tag{3.2}$$

where $\beta$ and $\gamma$ and constants of O(1), and the argument of the logarithm is to be evaluated in the limit $a \to A$. Assuming one can justify a cut-off in the logarithmic divergence at physical meaningful values (e.g. an uncertainty in a(t) corresponding to a Planck time uncertainty in t), then to within a couple of orders of magnitude

$$n_\phi \sim C^3 \sim t^3 \tag{3.3}$$

where I have evaluated the $1/a^3$ factor in Eq. (3.1) shortly after $a = A$, and used Eq. (2.9) for $a > A$ to substitute for $C$, assuming $\alpha$ is O(1). Equation (3.3) is far less dramatic, corresponding roughly to a few particles created per particle horizon volume. This means for $t \gg t_{\text{Planck}}$, $n_\phi \ll$ density of pre-existing particles. The stronger divergence found from using solution (2.4), which suggests a universe saturated with created particles, seems to be an artefact of the simplistic tachyon-only solution.

It is instructive to examine the change in the scalar field energy density around the abrupt tachyon disappearance. This is given by (Birrell & Davies 1982, Eq. (6.166))

$$\rho(t_2)a(t_2) - \rho(t_1)a(t_1) = -\int_{t_1}^{t_2} a^3 \langle T_\nu^\nu \rangle \, dt \tag{3.4}$$

where $\langle T_\nu^\nu \rangle$ is the trace of the expectation value of the stress-energy-momentum tensor of the scalar field evaluated in the conformal vacuum. The right hand side of Eq. (3.4) is proportional to $(\xi - 1/6)^2 R^2$ and so will diverge if Eq. (2.9) is used to evaluate $R$, with $\xi \neq 1/6$. If a smearing assumption is made then, to first approximation, the energy density of created particles will be given by $n_\phi \omega$, where Eq. (3.2) is used for $n_\phi$ and $\omega$ is a characteristic frequency given by the inverse of the smearing time scale. Of greater interest is the fact that the right hand side of Eq. (3.4) is non-zero even in the case of conformal coupling, $\xi = 1/6$, on account of the conformal trace anomaly. Indeed, the integral in Eq. (3.4) may be performed explicitly (Birrell & Davies 1982, Eq. (6.171)) to give

$$3Ea'^4 + 12F(-a^2 a' a''' - a a'^2 a'' + a^2 a''^2/2 + 3a'^4/2) \tag{3.5}$$

where the numerical coefficients are $E = (2880\pi^2)^{-1}$ and $F = -(17280\pi^2)^{-1}$. This expression diverges like $(1 - a^2/A^2)^{-3/2}$ as $a \to A$, representing an intense surge of vacuum energy, even though in this case there is no particle production. (The peak value of this energy will be determined by the form of the smearing function assumed for A.) Once the tachyons have disappeared, this vacuum energy falls back to a negligible value ~ $t^{-2}$.

## 4. Event horizon behaviour

A realistic cosmological model would involve non-zero dark energy in addition to the assumed tachyons and conventional matter and radiation. If a cosmological constant term is introduced into the Friedmann equations (2.1) and (2.2), the possibility arises of a cosmological event horizon. To investigate what effect the tachyons have on the dynamics of the horizon, I shall consider here a simplified model in which normal matter and radiation are absent. A comprehensive discussion of cosmological event horizons in the context of the generalised second law of thermodynamics has been given by Davies, Davis and Lineweaver (2003).

The radius of the horizon is defined for spatially flat FRW models by

$$R_h = a(t)\int_t^\infty dt/a(t) = a(t)\int_a^\infty da/aa' \qquad (4.1)$$

where the scale factor $a(t)$ is a solution of the Friedmann Eq. (2.1), augmented by a cosmological constant $\Lambda$:

$$a'^2/a^2 = \Lambda/3 + 8\pi G\rho_{tachyons} = \Lambda/3 + (\alpha/a^4)(1 - a^2/A^2)^{1/2} \qquad a < A$$
$$= \Lambda/3 \qquad a > A \qquad (4.2)$$

where again $\alpha$ is a constant that determines the total energy density of tachyons. Equation (4.2) cannot be integrated in terms of simple functions, so I shall consider the approximation $a \to A$. Taking the square root of Eq. (4.2) and substituting into Eq. (4.1) yields, in this approximation,

$$R_h \approx (3/\Lambda)^{1/2} - (3/\Lambda)^{3/2}(\alpha/2)\int_a^A a^{-6}(1 - a^2/A^2)^{1/2} da \qquad a < A$$
$$= (3/\Lambda)^{1/2} \qquad a > A. \qquad (4.3)$$

Performing the integral in Eq. (4.3), and putting a ≈ A, one obtains

$$R_h \approx (3/\Lambda)^{1/2} - (3/\Lambda)^{3/2}(\alpha/6)(1 - a^2/A^2)^{3/2} \qquad a < A$$
$$= (3/\Lambda)^{1/2} \qquad a > A. \qquad (4.4)$$

The term $(3/\Lambda)^{1/2}$ is the usual de Sitter horizon radius. The other term represents a reduction in horizon area caused by the tachyon fluid.

It follows that

$$R_h' \approx (\alpha/2A)(1 - a^2/A^2)^{1/2} \quad > 0, \qquad (4.5)$$

in conformity with the generalised second law of thermodynamics, in spite of the fact that tachyons themselves transcend this law by traversing event horizons. Perhaps more surprisingly, $R_h' \to 0$ smoothly as $a \to A$, even though the pressure diverges there.

## 5. Conclusions and observable consequences

Although tachyons remain extremely conjectural, the analysis presented here reveals no unphysical or pathological consequences for cosmological theory if some component of dark matter is in the form of a uniform tachyonic fluid. Even the generalised second law of thermodynamics applied to cosmological horizons remains valid. The central result – that tachyons disappear abruptly due to the expansion of the universe – probably limits their putative influence to the early or very early universe.

Can the presence of a tachyonic background lead to observable cosmological effects? In this paper I have limited my attention to the abrupt (formally divergent) behaviour of $a''$ and $R$ caused by the sudden disappearance of the tachyons. I showed that this discontinuity is associated with quantum particle production, but that the density of created scalar particles was very low as long as the tachyonic fluid is considered homogeneous and isotropic. It seems likely, however, that irregularities in the tachyon background would lead to local 'hot spots' of particle creation that might have observable consequences. Stronger effects were found to be associated with the vacuum energy, even in the case of conformal symmetry as might apply to, for example, to photons. But this effect is transient, limited to the epoch of tachyon disappearance.

I have not considered here the classical effects of a diverging $R$, such as the impact on gravitational density perturbations, or the back reaction of the tachyon fluid's rapidly changing energy and pressure on the cosmological dynamics. In particular, it is not obvious whether the gravitational back reaction will serve to homogenize or inhomogenize the tachyonic fluid. If $\mu \sim$ electron mass, as might seem reasonable on symmetry grounds, then $a = A$ at about 1s. If $\sim 50$ per cent of the energy density of the universe evaporates in the first one second of expansion in a phase transition involving escalating values of $p$ and $R$, possibly inhomogeneously distributed, the effects on the density spectrum of normal matter and radiation could be important. It is possible that the tachyons would imprint a distinctive fluctuation spectrum on the universe that might be detectable in future cosmic microwave background measurements.

A major uncertainty in this analysis is the energy distribution of the tachyons, which would have the effect of smearing the abrupt behaviour near $a = A$. In the absence of a proper theory of tachyon interactions, or a defensible ansatz about the tachyons' origin or initial energy distribution, this aspect must remain conjectural.

Probably the most important consequence of a tachyonic component of dark matter is the effect on the dynamics of the early universe, with concomitant implications for primordial nucleosynthesis. The relationship between temperature and time for a standard radiation-dominated FRW model universe is

$$T \propto (N)^{1/4} t^{-1/2} \tag{5.1}$$

where *N* now denotes the total number of all relativistic particle species. If half the particle species in the early universe were tachyonic, then the temperature would be suppressed by a factor 1.19 relative to a model with no tachyons.

Finally, I should point out that in this paper I have treated the tachyons in a standard manner as a classical background fluid. Recently, results from string theory have suggested the possible existence of quantum fields with tachyonic terms in the Lagrangian. Some attention has been given to possible cosmological consequences of such a field (Gibbons 2003; Bagla et. al. 2003). There does not seem to be any direct connection between this work and the foregoing.